\documentclass[aps,prc,twocolumn,groupedaddress,amsmath,amssymb,floatfix,superscriptaddress]{revtex4}
\usepackage{multirow}
\usepackage{graphicx}
\usepackage{color}
\usepackage{bm}
\usepackage{comment}
\usepackage{tablefootnote}
\usepackage{threeparttable}
\usepackage{hyperref}

\newcommand{\PerTonDay}{(ton\,day)$^{-1}$}
\newcommand{\PerTonYr}{(ton\,yr)$^{-1}$}

\begin{document}

\bibliographystyle{h-physrev3}

\title{Search for Majorana Neutrinos with the Complete KamLAND-Zen Dataset}

\newcommand{\tohoku}{\affiliation{Research Center for Neutrino
    Science, Tohoku University, Sendai 980-8578, Japan}}
\newcommand{\ipmu}{\affiliation{Kavli Institute for the Physics and Mathematics of the Universe (WPI), 
    The University of Tokyo Institutes for Advanced Study, 
    The University of Tokyo, Kashiwa, Chiba 277-8583, Japan}}
\newcommand{\osaka}{\affiliation{Graduate School of 
    Science, Osaka University, Toyonaka, Osaka 560-0043, Japan}}
\newcommand{\rcnp}{\affiliation{Research Center for Nuclear Physics, 
    Osaka University, Ibaraki, Osaka 567-0047, Japan}}
\newcommand{\tokushima}{\affiliation{Department of Physics, 
    Tokushima University, Tokushima 770-8506, Japan}}
\newcommand{\tokushimags}{\affiliation{Graduate School of Integrated Arts and Sciences, 
    Tokushima University, Tokushima 770-8502, Japan}}
\newcommand{\lbl}{\affiliation{Nuclear Science Division, Lawrence Berkeley National Laboratory,
    Berkeley, California 94720, USA}}
\newcommand{\hawaii}{\affiliation{Department of Physics and Astronomy,
    University of Hawaii at Manoa, Honolulu, Hawaii 96822, USA}}
\newcommand{\mitech}{\affiliation{Massachusetts Institute of Technology, 
    Cambridge, Massachusetts 02139, USA}}
\newcommand{\ut}{\affiliation{Department of Physics and
    Astronomy, University of Tennessee, Knoxville, Tennessee 37996, USA}}
\newcommand{\tunl}{\affiliation{Triangle Universities Nuclear Laboratory, Durham, 
    North Carolina 27708, USA; \\
    Physics Departments at Duke University, Durham, North Carolina 27708, USA; \\
    North Carolina Central University, Durham, North Carolina 27707, USA; \\
    and The University of North Carolina at Chapel Hill, Chapel Hill, North Carolina 27599, USA}}
\newcommand{\vt}{\affiliation{Center for Neutrino
   Physics, Virginia Polytechnic Institute and State University, Blacksburg,
   Virginia 24061, USA}}
\newcommand{\washington}{\affiliation{Center for Experimental Nuclear Physics and Astrophysics, 
    University of Washington, Seattle, Washington 98195, USA}}
\newcommand{\nikhef}{\affiliation{Nikhef and the University of Amsterdam, 
    Science Park, Amsterdam, the Netherlands}}
\newcommand{\gppu}{\affiliation{Graduate Program on Physics for the Universe, Tohoku University, Sendai 980-8578, Japan}}
\newcommand{\bu}{\affiliation{Boston University, Boston, Massachusetts 02215, USA}}
\newcommand{\chapel}{\affiliation{UNC Physics and Astronomy, 120 E. Cameron Ave., Phillips Hall CB3255, Chapel Hill, NC 27599}}
\newcommand{\obihiro}{\affiliation{Department of Human Science, Obihiro University of Agriculture and Veterinary Medicine, Obihiro, Hokkaido 080-8555, Japan}}
\newcommand{\sandiego}{\affiliation{Hal{\i}c{\i}o\u{g}lu Data Science Institute, Department of Physics, University of California San Diego, La Jolla, California, 92093, USA}}
\newcommand{\delaware}{\affiliation{Department of Physics and Astronomy, University of Delaware, Newark, Delaware 19716, USA}}

\newcommand{\aticrrnow}{\altaffiliation
    {Present address: Kamioka Observatory, Institute for Cosmic-Ray Research, 
    The University of Tokyo, Hida, Gifu 506-1205, Japan}}
\newcommand{\atrikennow}{\altaffiliation
    {Present address: Center for Advanced Photonics, 
    RIKEN, Wako, Saitama, 351-0198, Japan}}
\newcommand{\atbutsuryonow}{\altaffiliation
    {Present address: Faculty of Health Sciences, 
    Butsuryo College of Osaka, Sakai, Osaka 593-8328, Japan}}
\newcommand{\atkandainow}{\altaffiliation
    {Faculty of Environmental and Urban Engineering,
    Kansai University, Suita, Osaka 564-8680, Japan}}
\newcommand{\atkinkennow}{\altaffiliation
    {Present address: Institute for Materials Research, 
    Tohoku University, 2-1-1 Katahira, Aoba-ku, Sendai, 
    Miyagi, 980-8577, Japan}}

%
%
\author{S.~Abe}\aticrrnow\tohoku
\author{T.~Araki}\tohoku
\author{K.~Chiba}\tohoku
\author{T.~Eda}\tohoku
\author{M.~Eizuka}\tohoku
\author{Y.~Funahashi}\tohoku
\author{A.~Furuto}\tohoku
\author{A.~Gando}\tohoku
\author{Y.~Gando}\tohoku\obihiro
\author{S.~Goto}\tohoku
\author{T.~Hachiya}\tohoku
\author{K.~Hata}\tohoku
\author{K.~Ichimura}\tohoku
\author{S.~Ieki}\tohoku
\author{H.~Ikeda}\tohoku
\author{K.~Inoue}\tohoku
\author{K.~Ishidoshiro}\tohoku
\author{Y.~Kamei}\atrikennow\tohoku
\author{N.~Kawada}\tohoku
\author{Y.~Kishimoto}\tohoku
\author{M.~Koga}\tohoku\ipmu
\author{A.~Marthe}\tohoku
\author{Y.~Matsumoto}\tohoku
\author{T.~Mitsui}\atbutsuryonow\tohoku
\author{H.~Miyake}\tohoku\gppu
\author{D.~Morita}\tohoku
\author{R.~Nakajima}\tohoku
\author{K.~Nakamura}\atkandainow\tohoku
\author{R.~Nakamura}\tohoku
\author{R.~Nakamura}\tohoku
\author{J.~Nakane}\tohoku
\author{T.~Ono}\tohoku
\author{H.~Ozaki}\tohoku
\author{K.~Saito}\tohoku
\author{T.~Sakai}\tohoku
\author{I.~Shimizu}\tohoku
\author{J.~Shirai}\tohoku
\author{K.~Shiraishi}\tohoku
\author{A.~Suzuki}\tohoku
\author{K.~Tachibana}\tohoku
\author{K.~Tamae}\tohoku
\author{H.~Watanabe}\tohoku
\author{K.~Watanabe}\tohoku

\author{S.~Yoshida}\osaka

\author{S.~Umehara}\rcnp

\author{K.~Fushimi}\tokushima
\author{K.~Kotera}\tokushimags
\author{Y.~Urano}\atkinkennow\tokushimags

\author{B.E.~Berger}\lbl
\author{B.K.~Fujikawa}\ipmu\lbl

\author{J.G.~Learned}\hawaii
\author{J.~Maricic}\hawaii

\author{Z.~Fu}\mitech
\author{S.~Ghosh}\mitech
\author{J.~Smolsky}\mitech
\author{L.A.~Winslow}\mitech

\author{Y.~Efremenko}\ipmu\ut

\author{H.J.~Karwowski}\tunl
\author{D.M.~Markoff}\tunl
\author{W.~Tornow}\ipmu\tunl

\author{S.~Dell'Oro}\vt
\author{T.~O'Donnell}\vt

\author{J.A.~Detwiler}\ipmu\washington
\author{S.~Enomoto}\ipmu\washington

\author{M.P.~Decowski}\nikhef
\author{K.M.~Weerman}\nikhef

\author{C.~Grant}\bu
\author{\"{O}.~Penek}\bu
\author{H.~Song}\bu

\author{A.~Li}\sandiego

\author{S.N.~Axani}\delaware
\author{M.~Garcia}\delaware
\author{M.~Sarfraz}\delaware

\collaboration{KamLAND-Zen Collaboration}\noaffiliation

\date{\today}

\begin{abstract}
We present a search for neutrinoless double-beta ($0\nu\beta\beta$) decay of $^{136}$Xe using the full \mbox{KamLAND-Zen 800} dataset with 745\,kg of enriched xenon, corresponding to an exposure of $2.1$\,ton\,yr of $^{136}$Xe. This updated search benefits from a more than twofold increase in exposure, recovery of photo-sensor gain, and reduced background from muon-induced spallation of xenon. Combining with the search in the previous KamLAND-Zen phase, we obtain a lower limit for the $0\nu\beta\beta$ decay half-life of $T_{1/2}^{0\nu} > 3.8 \times 10^{26}$\,yr at 90\% C.L., a factor of 1.7 improvement over the previous limit. The corresponding upper limits on the effective Majorana neutrino mass are in the range 28--122\,meV using phenomenological nuclear matrix element calculations.
\end{abstract}

\maketitle

The search for neutrinoless double-beta ($0\nu\beta\beta$) decay is an active field of research in nuclear and particle physics~\cite{Agostini2023}. Observation of this decay would prove the Majorana nature of neutrinos and demonstrate the non-conservation of lepton number, a key feature of Leptogenesis, which can explain the imbalance of matter over antimatter in the Universe~\cite{Fukugita1986}. In the model of light Majorana neutrino exchange, the $0\nu\beta\beta$ decay rate is proportional to the square of the effective Majorana neutrino mass $\left<m_{\beta\beta}\right> \equiv \left| \Sigma_{i} U_{ei}^{2}m_{\nu_{i}} \right|$. The most stringent limit to date has been provided by KamLAND-Zen, reaching the inverted mass ordering (IO) region below 50\,meV for the first time~\cite{Abe2023a}. An improved search covering the unexplored IO region would stringently test theoretical models predicting $\left<m_{\beta\beta}\right>$ in this range~\cite{Harigaya2012,Asaka2020,Asai2020}, with a significantly increased chance to observe $0\nu\beta\beta$ decay~\cite{Agostini2017}.

The KamLAND-Zen (KamLAND Zero-Neutrino Double-Beta Decay)~\cite{Gando2012a,Gando2012b,Gando2013a,Asakura2016,Gando2016,Gando2019,Abe2023a} experiment utilizes the low-background KamLAND detector to search for $0\nu\beta\beta$ decay. Enriched xenon is dissolved in liquid scintillator (Xe-LS) contained in a teardrop-shaped inner balloon (IB). The IB is constructed from 25-$\mu$m-thick transparent nylon film and is suspended at the center of the KamLAND detector. The IB is surrounded by 1\,kton of LS (Outer LS) contained in a 13-m-diameter spherical outer balloon. Scintillation light is viewed by 1,325 17-inch and 554 20-inch photomultiplier tubes (PMTs) mounted on an 18-m-diameter spherical stainless-steel tank. In the previous phase of the experiment (\mbox{KamLAND-Zen 400}) with 381\,kg of enriched xenon, we probed the quasidegenerate neutrino mass region~\cite{Gando2016}. To further improve the search sensitivity, the \mbox{KamLAND-Zen} collaboration upgraded the experiment (\mbox{KamLAND-Zen 800}) to 745\,kg of enriched xenon, nearly twice the target mass of the previous phase. To hold the increased xenon, we constructed a larger and cleaner 3.80-m-diameter IB~\cite{Gando2021}. The \mbox{Xe-LS} consists of 82\% decane and 18\% pseudocumene (1,2,4-trimethylbenzene) by volume, 2.4\,g/liter of the fluor PPO (2,5-diphenyloxazole), and $(3.13 \pm 0.01)$\% by weight of isotopically enriched xenon gas. The isotopic abundances in the enriched xenon are $(90.85 \pm 0.13)\%$ \mbox{$^{136}$Xe}, $(8.82 \pm 0.01)\%$ \mbox{$^{134}$Xe}, $(0.17 \pm 0.01)\%$ \mbox{$^{132}$Xe}, and the remaining $0.16\%$ from others.

\mbox{KamLAND-Zen 800} completed data-taking in January 2024 to start dismantling the current KamLAND detector in preparation for a detector upgrade, \mbox{KamLAND2-Zen}. This Letter reports on the $0\nu\beta\beta$ decay analysis with the full KamLAND-Zen 800 dataset collected between February 5, 2019, and January 12, 2024, and its combination with the earlier KamLAND-Zen 400 data.

The experimental signature is a characteristic peak in the energy distribution, produced by the two $e^{-}$ emitted from $^{136}$Xe $\beta\beta$ decay. For hypothetical $0\nu\beta\beta$ decays, their summed energy is always 2.458\,MeV (\mbox{Q-value} of $^{136}$Xe $\beta\beta$ decay)~\cite{Redshaw2007}, while for $2\nu\beta\beta$ decays the sum has a continuous spectrum up to the \mbox{Q-value}. The event vertex and energy are reconstructed based on the timing and charge distributions of photoelectrons (p.e.) recorded by the PMTs. Energy calibration is performed using $^{214}$Bi $\beta$'s and $\gamma$'s from $^{222}$Rn (\mbox{$\tau = 5.5$\,days}), introduced during the initial filling of the IB with Xe-LS, and 2.225\,MeV $\gamma$'s from capture of spallation neutrons by protons. The detector Monte Carlo (MC) simulation is based on \texttt{Geant4}~\cite{Agostinelli2003,Allison2006} and is tuned to reproduce the timing and charge distributions observed in the data. The estimated energy and vertex resolutions in the \mbox{Xe-LS} are 6.7\%/$\sqrt{E({\rm MeV})}$ and 13.7\,cm/$\sqrt{E({\rm MeV})}$, respectively.

During the earlier period in KamLAND-Zen 800, we observed an increasing trend in the number of low photo-detection efficiency PMTs caused by age-related gain decrease. The resulting degradation in energy resolution could increase the background from the high energy tail of $2\nu\beta\beta$ events. In March 2020, we started a continuous campaign modifying the input circuitry of the KamLAND front-end electronics (FEE) to counter the PMT gain decrease. The 1 p.e.~gain and timing offsets due to the circuit modification are corrected for each PMT based on photon counting with high-statistics background events. Owing to these restoration efforts, we have succeeded in preventing the degradation of the energy resolution by a factor of 1.2, which would have led to a twofold increase of the $2\nu\beta\beta$ background.

\begin{figure}[t]
\begin{center}
\includegraphics[width=1.0\columnwidth]{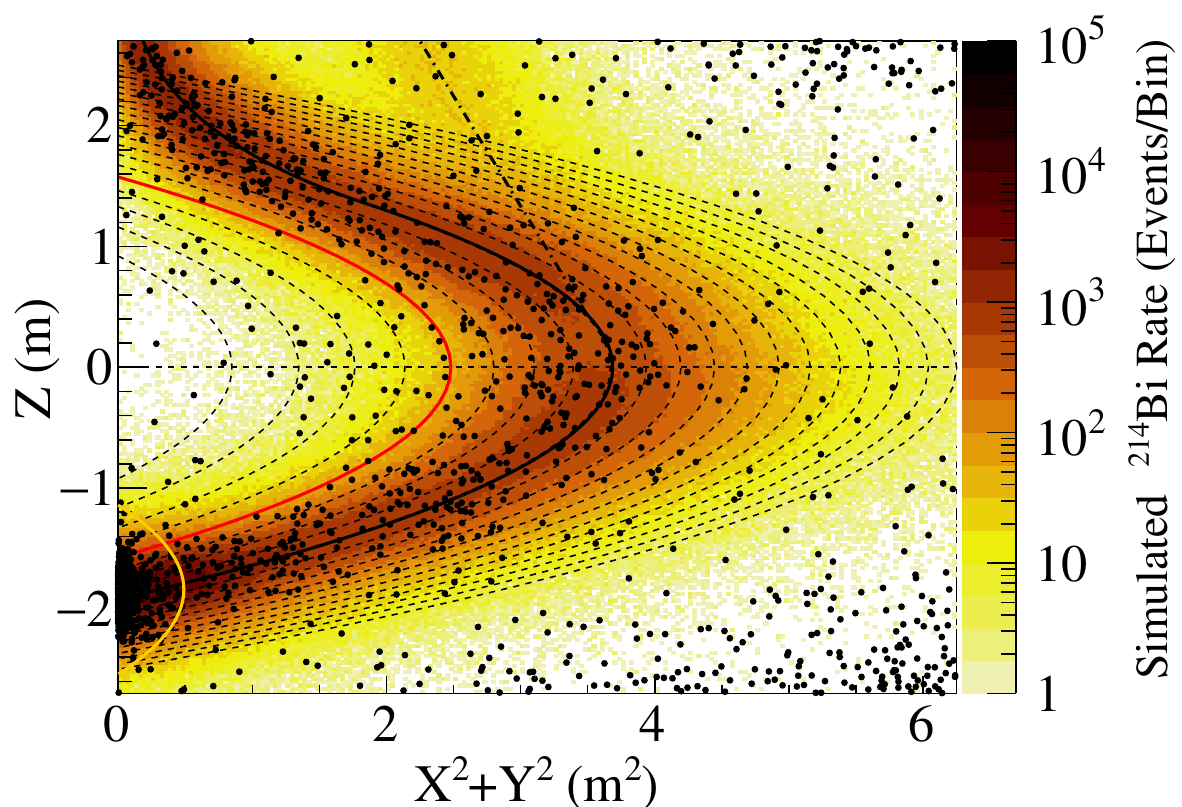}
\vspace{-0.8cm}
\end{center}
\caption{Vertex distribution of candidate singles data (SD) events (black points) in the $0\nu\beta\beta$ energy window $2.35 < E < 2.70\,{\rm MeV}$. The thick black and red lines indicate the shape of the IB and the 1.57-m-radius spherical volume, respectively. The dot-dashed line indicates the nylon belt suspending the IB. The thin dashed lines illustrate the shape of the equal-volume spherical half-shells, which compose the 2.5-m-radius spherical fiducial volume. The high-count region at the IB bottom within the yellow solid line indicates the hot spot and is removed through a 0.7\,m radius cut.}
\vspace{-0.3cm}
\label{figure:vertex}
\end{figure}

We apply the following cuts to select $\beta\beta$ decay events: (i) The reconstructed vertex must be within 2.5\,m of the detector center and 0.7\,m away from the bottom hot spot on the IB, which is outlined in Fig.~\ref{figure:vertex}. (ii) Muons and events within 2\,ms after muons are rejected. (iii) Bi-Po decay pairs are eliminated by a delayed coincidence tag, which requires the time and distance between the prompt Bi and delayed Po decay candidates to be less than 1.9\,ms and 1.7\,m, respectively, and by rejecting events with a double pulse identification within a single event acquisition window. Those cuts remove $(99.95 \pm 0.03)$\% of $^{214}$Bi-$^{214}$Po events, and $(97.9 \pm 0.5)$\% of $^{212}$Bi-$^{212}$Po. (iv) Reactor $\overline{\nu}_{e}$'s identified by a delayed coincidence of positrons and neutron-capture $\gamma$'s are rejected. (v) Poorly reconstructed events are rejected. These events are tagged using a vertex-time-charge discriminator which measures how well the observed PMT time-charge distributions match those expected based on the reconstructed vertex. The overall selection inefficiency for $0\nu\beta\beta$ events is less than 0.1\%.

Background sources external to the \mbox{Xe-LS} are dominated by radioactive impurities (RI) on the IB film. The contamination levels of $^{238}$U and $^{232}$Th are $(4.3 \pm 0.2)\times 10^{-12}$\,g/g and $(2.4 \pm 0.1)\times 10^{-11}$\,g/g, respectively, which are roughly a factor of 10 smaller than those measured on the previous IB~\cite{Gando2016}. The reference calculations for $^{238}$U and $^{232}$Th mentioned here assume secular equilibrium for comparison with the previously reported values. In the earlier period of the dataset, we found an increase in the background rate at the IB bottom, possibly due to the settling of dust particles containing radioactive impurities. To avoid this possible background, we tag and remove this high-background period from the dataset using machine learning algorithms, as discussed in Ref.~\cite{Abe2023a,Li2023}.

Solar neutrinos are an intrinsic background source for KamLAND-Zen. Based on  neutrino flux calculations using the standard solar model~\cite{Serenelli2011}, the elastic scattering (ES) of $^{8}$B solar neutrinos on electrons in the Xe-LS is estimated to be $(4.9 \pm 0.2) \times 10^{-3} $\,\PerTonDay, including the effect of three-flavor neutrino oscillations. The charged-current (CC) interactions on $^{136}$Xe produce $e^{-}$, $^{136}$Cs, and $\gamma$'s from the excited states, and the subsequent decays of $^{136}$Cs (\mbox{$\tau=19.0$\,days}, \mbox{$Q = 2.548$\,MeV}) create a background peak around 2.0\,MeV in visible energy, mostly overlapping with  the resolution tail of $2\nu\beta\beta$ decays. The interaction rate is expected to be $(0.8 \pm 0.1)\times  10^{-3}$\,\PerTonDay~based on the cross section calculated in Ref.~\cite{Ejiri2014,Frekers2013}.

Cosmic-ray muons produce neutrons and radioactive isotopes through nuclear spallation of carbon and xenon, which decay by emitting $\beta$'s or $\gamma$'s. To remove these events, we apply cuts based on space and time correlation with muon and neutron capture $\gamma$ events. In the $0\nu\beta\beta$ window, decays of $^{10}$C (\mbox{$\tau = 27.8$\,s}, \mbox{$Q = 3.65$\,MeV}) and $^{6}$He~($\tau = 1.16\,{\rm s}$, $Q = 3.51\,{\rm MeV}$) dominate the muon spallation backgrounds. To reduce these short-lived backgrounds, we remove events within 150\,ms after muons, and events reconstructed within 1.6\,m of neutron vertices for 180\,s. In addition, we reject remaining background events by the ``shower'' likelihood method based on reconstructed muon shower profiles~\cite{Li2014,Li2015a,Li2015b,Zhang2016,Abe2023b}. The overall rejection efficiencies for $^{10}$C and $^{6}$He are $>$$99.7$\% and $(97.3 \pm 1.5)\%$, respectively. To reduce the $^{137}$Xe ($\tau = 5.5\,{\rm min}$, $Q = 4.17\,{\rm MeV}$) background, we remove events reconstructed within 1.6\,m of vertices identified as neutron captures on ${}^{136}$Xe, which produce high energy $\gamma$’s ($Q = 4.03$~MeV), for 27 min. This cut removes $(74 \pm 7)\%$ of $^{137}$Xe. The deadtime introduced by these cuts is $(14.6 \pm 0.1)\%$.

The products of muon spallation with lifetimes greater than $\mathcal{O}$(100\,s), denoted as “long-lived products”, are attributed to the decay of heavy isotopes produced by xenon spallation. Xenon spallation can be tagged by detecting multiple neutrons. To characterize the long-lived products, we define a likelihood ratio, $R_{L} = L_{\rm spa} / (L_{\rm spa} + L_{\rm acc})$. Here $L_{\rm spa}$ and $L_{\rm acc}$ are the probability density functions (PDFs) for long-lived muon-spallation pairs and accidental pairs, respectively. These PDFs are constructed as a function of neutron multiplicity, distance to neutron vertices, and time interval from preceding muons. The cut value on $R_{L}$ is optimized using an estimate with MC simulation tools. \texttt{FLUKA}~\cite{BOHLEN2014211, Ferrari:898301} is used to calculate the spallation isotope yields, and \texttt{Geant4} to calculate the radioactive decays and energy spectra of the isotopes, considering their sequential decay chain. The predicted backgrounds are primarily from $^{132}$I, $^{130}$I, $^{124}$I, $^{122}$I, $^{118}$Sb, $^{110}$In, $^{88}$Y, and from other isotopes in smaller amounts (Table IX in Ref.~\cite{Abe2023b}). The total background rate in the $0\nu\beta\beta$ window is $30.0\pm2.2$\,\PerTonYr~in total, in units of $^{136}$Xe exposure. Events that are not classified as coming from long-lived backgrounds are referred to as ``singles data'' (SD), and the others are referred to as ``long-lived data'' (LD).

\begin{figure}
\includegraphics[width=1.0\columnwidth]{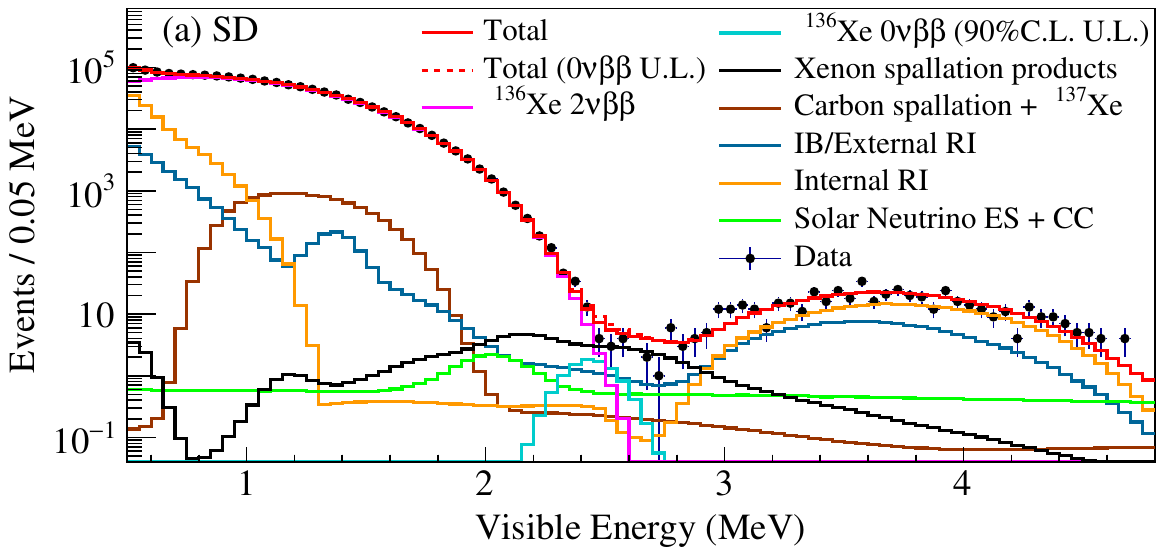}
\includegraphics[width=1.0\columnwidth]{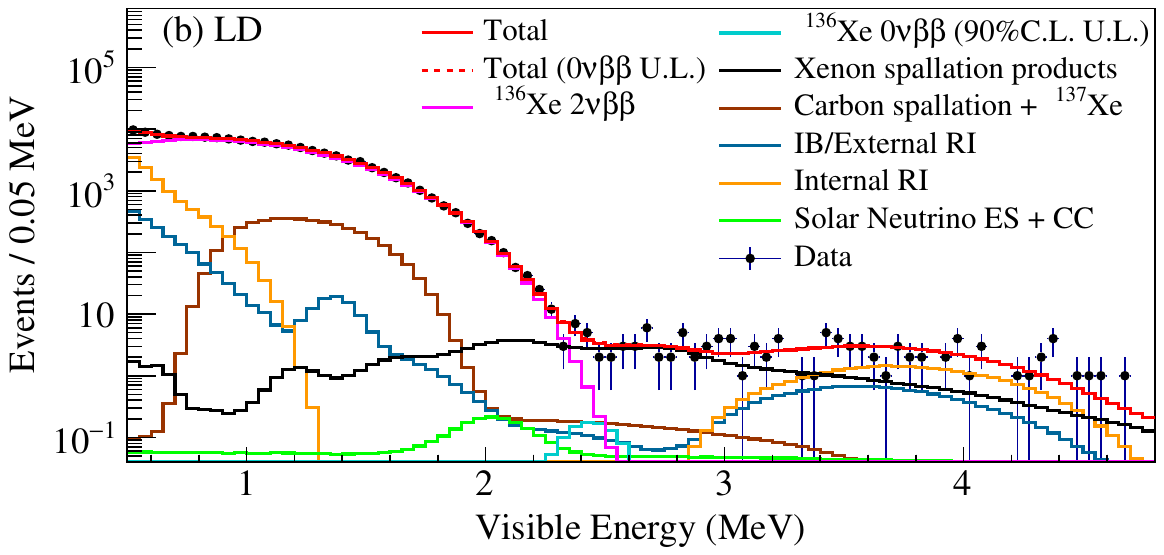}
\caption{Energy spectra of selected $\beta\beta$ candidates within a 1.57-m-radius spherical volume drawn together with best-fit backgrounds, the $2\nu\beta\beta$ decay spectrum, and the 90\% C.L. upper limit for $0\nu\beta\beta$ decay of (a) singles data (SD), and (b) long-lived data (LD). The LD exposure is about 10\% of the SD exposure.}
\label{figure:energy}
\end{figure}

We improved the neutron identification for post-muon events to enhance the tagging efficiency of the long-lived backgrounds. We identify these neutrons with a second, deadtime-free electronics system, called MoGURA~\cite{Abe2023b}. Previously, the neutron vertex was reconstructed using only PMTs with normal gain. In this analysis, we added low gain PMTs to the photon counting, achieving an increase in the mean number of PMT hits from $181$ to $201$ for 2.225\,MeV $\gamma$'s from neutron capture, after subtracting afterpulse noise contributions. The resulting neutron tagging efficiency is estimated to be $(74.5\pm0.4)\%$ based on the neutron capture time distribution, approximately 2\% better than in the previous analysis. Our MC study shows that $(44.0 \pm 8.7)\%$ of long-lived spallation backgrounds are classified as LD, whereas only $8.9$\% of uncorrelated events are mis-classified.

The total livetime for SD and LD in \mbox{KamLAND-Zen} 800 is $1133$\,days and $111$\,days, respectively. This is the effective measurement time considering the space and time cuts described in the background section. The total exposure of SD, which is sensitive to the $0\nu\beta\beta$ signal, is $2.1$\,ton\,yr of $^{136}$Xe, more than double the amount of the previous search~\cite{Abe2023a}. The $0\nu\beta\beta$ decay rate is estimated from a simultaneous likelihood fit (Statistics section in Ref.~\cite{PDG2024}) to the binned energy spectra of SD and LD between 0.5 and 4.8\,MeV, where systematic uncertainties are profiled in the likelihood. The binned energy spectra are made by dividing the 2.5-m-radius fiducial volume into 40 equal-volume bins, as illustrated in Fig.~\ref{figure:vertex}. The time variation of event rates is also considered in the fit. The contributions from major backgrounds in the \mbox{Xe-LS}, such as $^{85}$Kr, $^{40}$K, $^{210}$Bi, the $^{228}$Th-$^{208}$Pb subchain of the $^{232}$Th series, and long-lived spallation products, are free parameters and are left unconstrained in the spectral fit. The contributions from the $^{222}$Rn-$^{210}$Pb subchain of the $^{238}$U series and short-lived spallation products are constrained by their independent measurements~\cite{Abe2010}. The parameters of the detector energy response model are constrained by $^{214}$Bi data. For the long-lived spallation products an energy spectral distortion parameter based on linear rescaling is applied and constrained by its estimated uncertainty (Fig. 10 in Ref. [26]).

\begin{figure}
\vspace{-0.4cm}
\includegraphics[width=1.0\columnwidth]{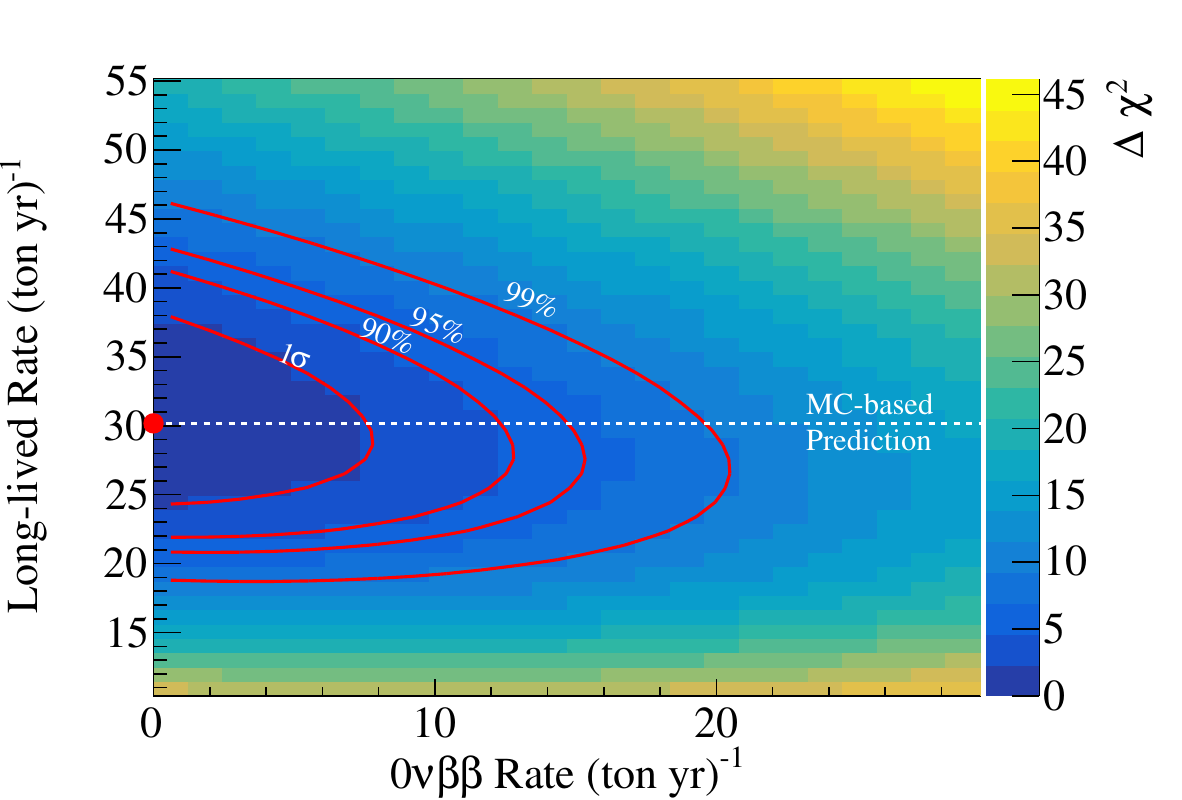}
\caption{Allowed region for the $^{136}$Xe $0\nu\beta\beta$ rate and the long-lived spallation background rate in the energy region $2.35 < E < 2.70\,{\rm MeV}$ ($0\nu\beta\beta$ window). The red dot and contour lines correspond to the best-fit and the $1\sigma$, 90\%, 95\%, 99\% C.L. intervals for two degrees of freedom, respectively. The horizontal line indicates the MC-based prediction.}
\vspace{-0.3cm}
\label{figure:chi2}
\end{figure}

Figure~\ref{figure:energy} shows the energy spectra of selected candidate SD and LD events within a 1.57-m-radius spherical volume together with the best-fit curves. The exposure of $^{136}$Xe for SD in this reduced volume is 1.13\,ton\,yr. The best-fit background contributions are summarized in Table~\ref{table:background}. The $0\nu\beta\beta$ signal best-fit is 0 events, indicating no event excess over the background expectation. Based on a profile likelihood method, we obtain a 90\% confidence level (C.L.) upper limit on the number of $0\nu\beta\beta$ decays of $<$\,$9.7$\,events ($<$\,$7.8$\,events in the range $2.35 < E < 2.70\,{\rm MeV}$), which corresponds to a limit of $<$\,$8.8$\,\PerTonYr~in units of $^{136}$Xe exposure, or $T_{1/2}^{0\nu\beta\beta} > 3.5 \times 10^{26}$\,yr (90\% C.L.). An analysis based on the Feldman-Cousins procedure~\cite{PhysRevD.57.3873_FCmethod} gives a slightly stronger limit of $4.3 \times 10^{26}$\,yr (90\% C.L.), indicating that the physical boundary at zero has only a limited impact on the $0\nu\beta\beta$ rate estimate, even with low statistics. An MC simulation of an ensemble of experiments assuming the best-fit background spectrum without a $0\nu\beta\beta$ signal indicates a median sensitivity of $2.2 \times 10^{26}$\,yr. The probability of obtaining a limit stronger than that reported here is 20\%. In addition to the frequentist analyses above, we also performed a statistical analysis within the Bayesian framework, assuming a flat prior for $1/T_{1/2}^{0\nu\beta\beta}$. The Bayesian limit and median sensitivity are $3.4 \times 10^{26}$\,yr and $2.4 \times 10^{26}$\,yr (90\% C.L.), respectively.

\begin{figure}[t]
\vspace{-0.3cm}
\includegraphics[width=1.0\columnwidth]{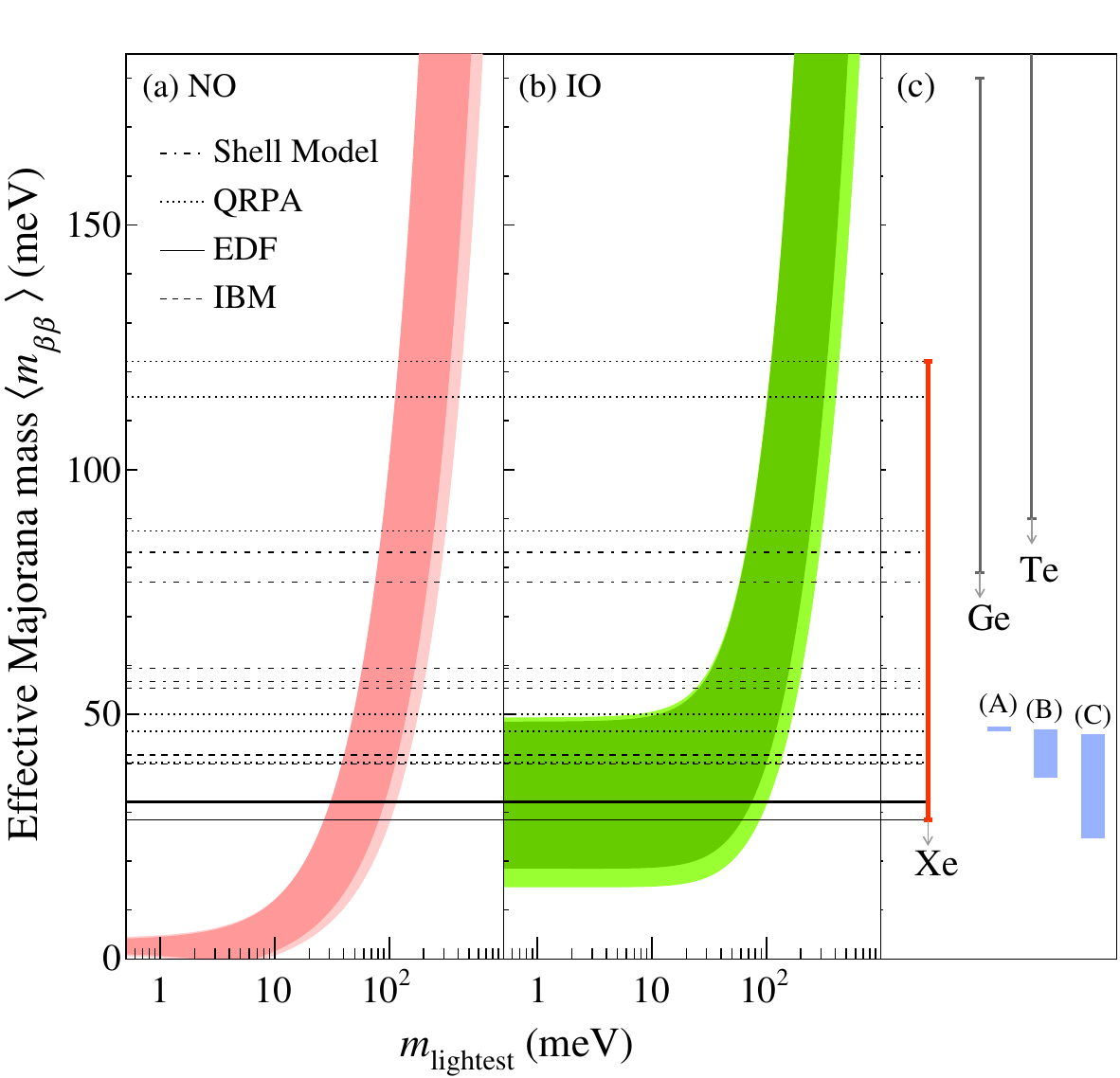}
\vspace{-0.3cm}
\caption{Effective Majorana neutrino mass $\left<m_{\beta\beta}\right>$ as a function of the lightest neutrino mass $m_{\rm lightest}$. The dark shaded regions are predictions based on best-fit values of neutrino oscillation parameters for (a) the normal ordering (NO) and (b) the inverted ordering (IO), and the light shaded regions indicate the $3\sigma$ ranges calculated from oscillation parameter uncertainties~\cite{DellOro2014,NuFIT2020}. The horizontal lines indicate 90\% C.L. upper limits on $\left<m_{\beta\beta}\right>$ with $^{136}$Xe from \mbox{KamLAND-Zen} (this work), considering an improved phase space factor calculation~\cite{Kotila2012,Stoica2013} and phenomenological NME calculations: shell model~\cite{Menendez2018, PhysRevC.93.024308horoi, PhysRevC.101.044315.coraggio2020, PhysRevC.105.034312.coraggio2022} (dot-dashed lines), quasiparticle randomphase approximation (QRPA)~\cite{PhysRevC.87.064302.mustonen, PhysRevC.91.024613.Hyv, PhysRevC.98.064325.simko2018, PhysRevC.97.045503fang, PhysRevC.102.044303terasaki}(dotted lines), energy-density functional (EDF) theory~\cite{PhysRevLett.105.252503.rodriguez, PhysRevLett.111.142501lopez, PhysRevC.95.024305song} (solid lines), interacting boson model (IBM)~\cite{PhysRevC.91.034304barea,  PhysRevD.102.095016deppisch} (dashed lines). (c) The corresponding limits for $^{136}$Xe, $^{76}$Ge~\cite{Agostini2020}, and $^{130}$Te~\cite{Adams2022}. Three theoretical predictions in the IO region are also shown: (A)\cite{Harigaya2012}, (B)\cite{Asaka2020}, (C)\cite{Asai2020}.}
\vspace{-0.3cm}
\label{figure:effective_mass}
\end{figure}

\begin{table}[t]
\caption[]{Summary of the estimated and best-fit background contributions in the energy region $2.35 < E < 2.70\,{\rm MeV}$ within the 1.57-m-radius spherical volume. In total, $60$ events were observed.}
\label{table:background}
\begin{threeparttable}[h]
\begin{tabular}{lcc}
\hline
\hline
Background \hspace{1.0cm} & ~~~~~~~Estimated~~~~~~~ & ~~~~~~Best-fit~~~~~~ \\
\hline
$^{136}$Xe $2\nu\beta\beta$ & $27.6 \pm 0.2$ \tnote{$\dagger$} & $27.68$ \\
\multicolumn{3}{c}{Residual radioactivity in \mbox{Xe-LS}} \\
\hline
$^{238}$U series & $0.08 \pm 0.01$ & $0.08$ \\
$^{232}$Th series & - & $1.35$ \\
\multicolumn{3}{c}{External (Radioactivity in IB)} \\
\hline
$^{238}$U series & - & $6.61$ \\
$^{232}$Th series & - & $0.04$ \\
\multicolumn{3}{c}{Neutrino interactions} \\
\hline
$^{8}$B solar $\nu$ $e^{-}$ ES & $3.57 \pm 0.09$ & $3.58$ \\
\multicolumn{3}{c}{Spallation products} \\
\hline
Long-lived & $17.82 \pm 1.30$ \tnote{$\dagger$} & $20.27$ \\
$^{10}$C & $0.00 \pm 0.05$ & $0.00$ \\
$^{6}$He & $0.54 \pm 0.27$ & $0.56$ \\
$^{137}$Xe & $0.72 \pm 0.60$ & $0.71$ \\
\hline
\hline
\end{tabular}
\begin{tablenotes}
\item[$\dagger$] Estimations based on the previous half-life measurements ($^{136}$Xe $2\nu\beta\beta$) and the spallation MC study (long-lived).
Those event rate constraints are not applied in the spectral fit.

\end{tablenotes}
\end{threeparttable}
\end{table}

Figure~\ref{figure:chi2} shows the allowed region of the $^{136}$Xe $0\nu\beta\beta$ rate and the long-lived spallation background rate from a combined fit of the KamLAND-Zen 400 and 800 datasets, giving a limit of 
\begin{equation}
    T_{1/2}^{0\nu\beta\beta} > 3.8 \times 10^{26}\,\textrm{yr~~~(90\% C.L.)}.
\end{equation}
The best-fit long-lived spallation background rate is $30.2^{+5.0}_{-4.0} $\,\PerTonYr indicating good consistency between the MC-based prediction and the LD analysis. This combined analysis has a sensitivity of $2.4 \times 10^{26}$\,yr, and the probability of obtaining a stronger limit is $20$\%. The lower limit on $T_{1/2}^{0\nu\beta\beta}$ can be converted to an upper limit on $\langle m_{\beta\beta}\rangle$ assuming that the $0\nu\beta\beta$ decay mechanism is dominated by exchange of light Majorana neutrinos. Assuming the axial coupling constant $g_{A} \simeq 1.27$, we obtain an upper limit of 
\begin{equation}
    \langle m_{\beta\beta}\rangle < (28-122)\,\textrm{meV~~~(90\% C.L.)},
    \label{eq:effmasslimit}
\end{equation}
using an improved phase space factor calculation of $14.54 \times 10^{-15}\, (\textrm{yr})^{-1}$~\cite{Kotila2012,Stoica2013}, and phenomenological nuclear matrix element (NME) calculations, summarized in Table~\ref{tab:NME}. In this letter, we follow the current standard to use those phenomenological models, although recent theoretical work has produced new NME calculations~\cite{belley2023ab, Jokiniemi2023} incorporating both quenching physics~\cite{Faessler_2008, PhysRevC.87.014315} and short-range contact~\cite{PhysRevLett.120.202001} contributions, which are missing in the calculations in Table~\ref{tab:NME}. Considering the calculations from Ref.~\cite{belley2023ab,Jokiniemi2023} gives ranges of upper limits on $\langle m_{\beta\beta}\rangle$ of $71-125$\,meV (ab initio~\cite{belley2023ab}), $25-60$\,meV (pnQRPA~\cite{Jokiniemi2023}), and $44-138$\,meV (shell model~\cite{Jokiniemi2023}). Figure \ref{figure:effective_mass} illustrates the allowed range of $\langle m_{\beta\beta} \rangle$ as a function of the lightest neutrino mass for the NME values in Table~\ref{tab:NME}. The shaded regions include the uncertainties in mixing elements $U_{ei}$ and the neutrino mass splitting for each hierarchy. Also drawn are experimental limits from $0\nu\beta\beta$ decay searches in other nuclei~\cite{Agostini2020,Adams2022}. The limit using $^{136}$Xe by \mbox{KamLAND-Zen} has provided the most stringent tests of the effective Majorana mass in the IO region and several theoretical models~\cite{Harigaya2012,Asaka2020,Asai2020}. It also provides the strongest constraint on $m_{\rm lightest}$ considering extreme cases of the combination of {\it CP} phases and the uncertainties from neutrino oscillation parameters~\cite{DellOro2014,NuFIT2020}. We obtain a 90\% C.L. upper limit of $m_{\rm lightest} < (84 - 353)\,{\rm meV}$. 

In conclusion, we presented a $0\nu\beta\beta$ decay search with the complete \mbox{KamLAND-Zen} dataset including improvements in the rejection of muon-induced long-lived spallation backgrounds. This search places the most stringent constraints to date on $\left<m_{\beta\beta}\right>$, and reaches the inverted mass ordering region below 50\,meV with half of the phenomenological nuclear matrix element calculations. The sensitivity is limited by background from xenon spallation products and $2\nu\beta\beta$ decay. In the near future, we plan to upgrade the detector (\mbox{KamLAND2-Zen}) with more ${}^{136}$Xe mass, enhanced energy resolution to minimize the leakage of the $2\nu\beta\beta$ tail into the $0\nu\beta\beta$ analysis window, and enhanced rejection of spallation products through improved neutron tagging and particle identification methods.

\begin{table}[b]
    \centering
    \caption[]{Summary of commonly used phenomenological nuclear matrix elements ($M^{0\nu}$) for $^{136}$Xe $0\nu\beta\beta$ decay and the corresponding 90\% C.L. upper limits on the effective Majorana neutrino mass $\left<m_{\beta\beta}\right>$ from KamLAND-Zen data. The limits shown in Eq.~(\ref{eq:effmasslimit}) use the minimum and maximum.} 
    \begin{tabular}{lccc}
        \hline\hline
        &  ~~~~~Ref. ~~~ &  ~~~$M^{0\nu}$ ~~~ &  ~~~$\langle m_{\beta\beta}\rangle$ (meV)~~~ \\
        \hline
        \multirow{3}{*}{Shell model}& \cite{Menendez2018} & 2.28, 2.45 & 59.4, 55.3 \\
                                    & \cite{PhysRevC.93.024308horoi} & 1.63, 1.76 & 83.1, 77.0\\
                                    & \cite{PhysRevC.101.044315.coraggio2020, PhysRevC.105.034312.coraggio2022} & 2.39 & 56.7\\
        \hline                             
        \multirow{5}{*}{QRPA}       & \cite{PhysRevC.87.064302.mustonen} & 1.55 & 87.4\\
                                    & \cite{PhysRevC.91.024613.Hyv} & 2.91 & 46.6\\
                                    & \cite{PhysRevC.98.064325.simko2018} & 2.71 & 50.0\\
                                    & \cite{PhysRevC.97.045503fang}  & 1.11, 1.18 & 122, 115\\
                                    & \cite{PhysRevC.102.044303terasaki} & 3.38 & 40.1\\
        \hline                                    
        \multirow{3}{*}{EDF theory} & \cite{PhysRevLett.105.252503.rodriguez} & 4.20 & 32.3\\
                                    & \cite{PhysRevLett.111.142501lopez} & 4.77 & 28.4\\
                                    & \cite{PhysRevC.95.024305song} & 4.24 & 32.0\\
        \hline                            
        \multirow{2}{*}{IBM}        & \cite{PhysRevC.91.034304barea} & 3.25 & 41.7\\
                                    & \cite{PhysRevD.102.095016deppisch} & 3.40 & 39.9\\
        \hline\hline        
    \end{tabular}
    \label{tab:NME}
\end{table}

\begin{acknowledgments}
The \mbox{KamLAND-Zen} experiment is supported by JSPS KAKENHI Grants No. 21000001, No. 26104002, and No. 19H05803; the U.S. National Science Foundation awards no. 2110720 and no. 2012964; the Heising-Simons Foundation; the Dutch Research Council (NWO); and under the U.S. Department of Energy (DOE) Grant No.\,DE-AC02-05CH11231, as well as other DOE and NSF grants to individual institutions. The Kamioka Mining and Smelting Company has provided service for activities in the mine. We acknowledge the support of NII for SINET4.
\end{acknowledgments}

\bibliography{DoubleBeta}

\begin{thebibliography}{10}

\bibitem{Agostini2023}
M.~Agostini, G.~Benato, J.~A. Detwiler, J.~Men{\'e}ndez, and F.~Vissani,
\newblock Rev. Mod. Phys. {\bf 95}, 025002 (2023).

\bibitem{Fukugita1986}
M.~Fukugita and T.~Yanagida,
\newblock Phys. Lett. B {\bf 174}, 45 (1986).

\bibitem{Abe2023a}
S.~Abe {\em et~al.}, (KamLAND-Zen Collaboration),
\newblock Phys. Rev. Lett. {\bf 130}, 051801 (2023).

\bibitem{Harigaya2012}
K.~Harigaya, M.~Ibe, and T.~T. Yanagida,
\newblock Phys. Rev. D {\bf 86}, 013002 (2012).

\bibitem{Asaka2020}
T.~Asaka, Y.~Heo, and T.~Yoshida,
\newblock Phys. Lett. B {\bf 811}, 135956 (2020).

\bibitem{Asai2020}
K.~Asai,
\newblock The European Physical Journal C {\bf 80}, 76 (2020).

\bibitem{Agostini2017}
M.~Agostini, G.~Benato, and J.~A. Detwiler,
\newblock Phys. Rev. D {\bf 96}, 053001 (2017).

\bibitem{Gando2012a}
A.~Gando {\em et~al.}, (KamLAND-Zen Collaboration),
\newblock Phys. Rev. C {\bf 85}, 045504 (2012).

\bibitem{Gando2012b}
A.~Gando {\em et~al.}, (KamLAND-Zen Collaboration),
\newblock Phys. Rev. C {\bf 86}, 021601 (2012).

\bibitem{Gando2013a}
A.~Gando {\em et~al.}, (KamLAND-Zen Collaboration),
\newblock Phys. Rev. Lett. {\bf 110}, 062502 (2013).

\bibitem{Asakura2016}
K.~Asakura {\em et~al.}, (KamLAND-Zen Collaboration),
\newblock Nucl. Phys. A {\bf 946}, 171 (2016).

\bibitem{Gando2016}
A.~Gando {\em et~al.}, (KamLAND-Zen Collaboration),
\newblock Phys. Rev. Lett. {\bf 117}, 082503 (2016).

\bibitem{Gando2019}
A.~Gando {\em et~al.}, (KamLAND-Zen Collaboration),
\newblock Phys. Rev. Lett. {\bf 122}, 192501 (2019).

\bibitem{Gando2021}
Y.~Gando {\em et~al.}, (KamLAND-Zen Collaboration),
\newblock JINST {\bf 16}, P08023 (2021).

\bibitem{Redshaw2007}
M.~Redshaw, E.~Wingfield, J.~McDaniel, and E.~G. Myers,
\newblock Phys. Rev. Lett. {\bf 98}, 053003 (2007).

\bibitem{Agostinelli2003}
S.~Agostinelli {\em et~al.},
\newblock Nucl. Instr. Meth. A {\bf 506}, 250  (2003).

\bibitem{Allison2006}
J.~Allison {\em et~al.},
\newblock IEEE Trans. Nucl. Sci. {\bf 53}, 270 (2006).

\bibitem{Li2023}
A.~Li {\em et~al.},
\newblock Phys. Rev. C {\bf 107}, 014323 (2023).

\bibitem{Serenelli2011}
A.~M. Serenelli, W.~C. Haxton, and C.~Pe{\~n}a-Garay,
\newblock Astrophys. J. {\bf 743}, 24 (2011).

\bibitem{Ejiri2014}
H.~Ejiri and S.~R. Elliott,
\newblock Phys. Rev. C {\bf 89}, 055501 (2014).

\bibitem{Frekers2013}
D.~Frekers, P.~Puppe, J.~H. Thies, and H.~Ejiri,
\newblock Nucl. Phys. A {\bf 916}, 219 (2013).

\bibitem{Li2014}
S.~W. Li and J.~F. Beacom,
\newblock Phys. Rev. C {\bf 89}, 045801 (2014).

\bibitem{Li2015a}
S.~W. Li and J.~F. Beacom,
\newblock Phys. Rev. D {\bf 91}, 105005 (2015).

\bibitem{Li2015b}
S.~W. Li and J.~F. Beacom,
\newblock Phys. Rev. D {\bf 92}, 105033 (2015).

\bibitem{Zhang2016}
Y.~Zhang {\em et~al.}, (Super-Kamiokande Collaboration),
\newblock Phys. Rev. D {\bf 93}, 012004 (2016).

\bibitem{Abe2023b}
S.~Abe {\em et~al.}, (KamLAND-Zen Collaboration),
\newblock Phys. Rev. C {\bf 107}, 054612 (2023).

\bibitem{BOHLEN2014211}
T.~B{\"o}hlen {\em et~al.},
\newblock Nuclear Data Sheets {\bf 120}, 211 (2014).

\bibitem{Ferrari:898301}
A.~Ferrari and \textit{et al.},
\newblock {\em {FLUKA: A multi-particle transport code (program version 2005) }}CERN Yellow Reports: Monographs (CERN, Geneva, 2005).

\bibitem{PDG2024}
S.~Navas {\em et~al.}, (Particle Data Group),
\newblock Phys. Rev. D {\bf 110}, 030001 (2024).

\bibitem{Abe2010}
S.~Abe {\em et~al.}, (KamLAND Collaboration),
\newblock Phys. Rev. C {\bf 81}, 025807 (2010).

\bibitem{PhysRevD.57.3873_FCmethod}
G.~J. Feldman and R.~D. Cousins,
\newblock Phys. Rev. D {\bf 57}, 3873 (1998).

\bibitem{DellOro2014}
S.~Dell'Oro, S.~Marcocci, and F.~Vissani,
\newblock Phys. Rev. D {\bf 90}, 033005 (2014).

\bibitem{NuFIT2020}
Nufit 5.0, available at http://www.nu-fit.org (2020).

\bibitem{Kotila2012}
J.~Kotila and F.~Iachello,
\newblock Phys. Rev. C {\bf 85}, 034316 (2012).

\bibitem{Stoica2013}
S.~Stoica and M.~Mirea,
\newblock Phys. Rev. C {\bf 88}, 037303 (2013); updated in arXiv:1411.5506v3 .

\bibitem{Menendez2018}
J.~Men{\'e}ndez,
\newblock J. of Phys. G {\bf 45}, 014003 (2018).

\bibitem{PhysRevC.93.024308horoi}
M.~Horoi and A.~Neacsu,
\newblock Phys. Rev. C {\bf 93}, 024308 (2016).

\bibitem{PhysRevC.101.044315.coraggio2020}
L.~Coraggio, A.~Gargano, N.~Itaco, R.~Mancino, and F.~Nowacki,
\newblock Phys. Rev. C {\bf 101}, 044315 (2020).

\bibitem{PhysRevC.105.034312.coraggio2022}
L.~Coraggio {\em et~al.},
\newblock Phys. Rev. C {\bf 105}, 034312 (2022).

\bibitem{PhysRevC.87.064302.mustonen}
M.~T. Mustonen and J.~Engel,
\newblock Phys. Rev. C {\bf 87}, 064302 (2013).

\bibitem{PhysRevC.91.024613.Hyv}
J.~Hyv\"arinen and J.~Suhonen,
\newblock Phys. Rev. C {\bf 91}, 024613 (2015).

\bibitem{PhysRevC.98.064325.simko2018}
F.~\u{S}imkovic, A.~Smetana, and P.~Vogel,
\newblock Phys. Rev. C {\bf 98}, 064325 (2018).

\bibitem{PhysRevC.97.045503fang}
D.-L. Fang, A.~Faessler, and F.~\u{S}imkovic,
\newblock Phys. Rev. C {\bf 97}, 045503 (2018).

\bibitem{PhysRevC.102.044303terasaki}
J.~Terasaki,
\newblock Phys. Rev. C {\bf 102}, 044303 (2020).

\bibitem{PhysRevLett.105.252503.rodriguez}
T.~R. Rodr\'{\i}guez and G.~Mart\'{\i}nez-Pinedo,
\newblock Phys. Rev. Lett. {\bf 105}, 252503 (2010).

\bibitem{PhysRevLett.111.142501lopez}
N.~L. Vaquero, T.~R. Rodr\'{\i}guez, and J.~L. Egido,
\newblock Phys. Rev. Lett. {\bf 111}, 142501 (2013).

\bibitem{PhysRevC.95.024305song}
L.~S. Song, J.~M. Yao, P.~Ring, and J.~Meng,
\newblock Phys. Rev. C {\bf 95}, 024305 (2017).

\bibitem{PhysRevC.91.034304barea}
J.~Barea, J.~Kotila, and F.~Iachello,
\newblock Phys. Rev. C {\bf 91}, 034304 (2015).

\bibitem{PhysRevD.102.095016deppisch}
F.~F. Deppisch, L.~Graf, F.~Iachello, and J.~Kotila,
\newblock Phys. Rev. D {\bf 102}, 095016 (2020).

\bibitem{Agostini2020}
M.~Agostini {\em et~al.}, (GERDA Collaboration),
\newblock Phys. Rev. Lett. {\bf 125}, 252502 (2020).

\bibitem{Adams2022}
D.~Q. Adams {\em et~al.}, (CUORE Collaboration),
\newblock Nature {\bf 604}, 53 (2022).

\bibitem{belley2023ab}
A.~Belley, T.~Miyagi, S.~R. Stroberg, and J.~D. Holt,
\newblock arXiv:2307.15156v1.

\bibitem{Jokiniemi2023}
L.~Jokiniemi, B.~Romeo, P.~Soriano, and J.~Men\'endez,
\newblock Phys. Rev. C {\bf 107}, 044305 (2023).

\bibitem{Faessler_2008}
A.~Faessler {\em et~al.},
\newblock J. Phys. G: Nucl. Part. Phys. {\bf 35}, 075104 (2008).

\bibitem{PhysRevC.87.014315}
J.~Barea, J.~Kotila, and F.~Iachello,
\newblock Phys. Rev. C {\bf 87}, 014315 (2013).

\bibitem{PhysRevLett.120.202001}
V.~Cirigliano {\em et~al.},
\newblock Phys. Rev. Lett. {\bf 120}, 202001 (2018).

\end{thebibliography}

\end{document}